\documentclass[a4paper]{jpconf}
\usepackage{graphicx}
\begin{document}
\title{The black hole information paradox and macroscopic superpositions}

\author{Stephen D.H. Hsu}

\address{Institute for Theoretical Science, University of Oregon, Eugene OR 97403-5203, USA}

\ead{hsu@uoregon.edu}

\begin{abstract}
We investigate the experimental capabilities required to test whether
black holes destroy information. We show that an experiment capable of
illuminating the information puzzle must necessarily be able to detect or
manipulate macroscopic superpositions (i.e., Everett branches). Hence,
it could also address the fundamental question of decoherence versus
wavefunction collapse.
\end{abstract}

\section{Introduction: black hole information}

In 1976 Hawking proposed that black holes destroy information: pure states which collapse to form black holes evaporate into mixed states described by density matrices \cite{HawkingPureMixed}. The argument in favor of information destruction can be pared to a few essential components; for reviews, see \cite{review1,review2,review2a,review2b,review3}. Hawking radiation, into which the hole evaporates, originates from outside the horizon and is causally disconnected from the interior: a spacelike slice can be constructed which intersects both the infalling matter and the outgoing Hawking radiation. The no-cloning theorem \cite{noclone} in quantum mechanics prevents information from residing in two places on the same slice, so the outgoing radiation must be independent of the initial state. (See Figure 1.)

It is safe to say that, over 30 years after Hawking's paper, theoreticians remain divided as to whether Hawking was originally correct, or whether some locality-violating mechanism somehow allows the information to escape. 

In this discussion \cite{original}, we investigate the following questions: Is the black hole information puzzle simply philosophy, or is it subject to experimental test? If the latter, then what capabilities are required? We find connections to a different question, from the foundations of quantum mechanics: do wavefunctions collapse, or is quantum evolution strictly unitary, leading inevitably to macroscopic superposition states? An experiment which sheds light on the black hole information puzzle would also be capable of addressing fundamental issues in quantum mechanics. See Zeh \cite{zeh} and also \cite{kiefernikolic} for related discussion.

To highlight the importance of macroscopic superpositions, we emphasize that the very formulation of the information puzzle relies on the use of the semiclassical black hole spacetime -- e.g., in the construction of the spacelike slice used in the no-cloning argument, or in the original Hawking calculation. But, in any fully quantum mechanical treatment of the black hole formation process there exist branches of the wavefunction, possibly of very small amplitude, in which no apparent horizon is formed and the initial state particles all escape to infinity. In other words, in which the spacetime is radically different. This is most easily seen if one considers black holes formed in the collision of two particles \cite{bhp} (see Figure \ref{bhf}): there is always a non-zero amplitude for no scattering -- i.e., no black hole creation -- in which the particles simply pass each other. This remains the case even as the number of particles in the initial state becomes large, although for certain semiclassical initial data one can make the no-formation amplitude arbitrarily small. Nevertheless, the information puzzle cannot avoid the issue of macroscopic superpositions: could such small amplitude branches restore purity or unitarity? \cite{scatter}

\begin{figure}
\begin{center}
\includegraphics[width=18pc]{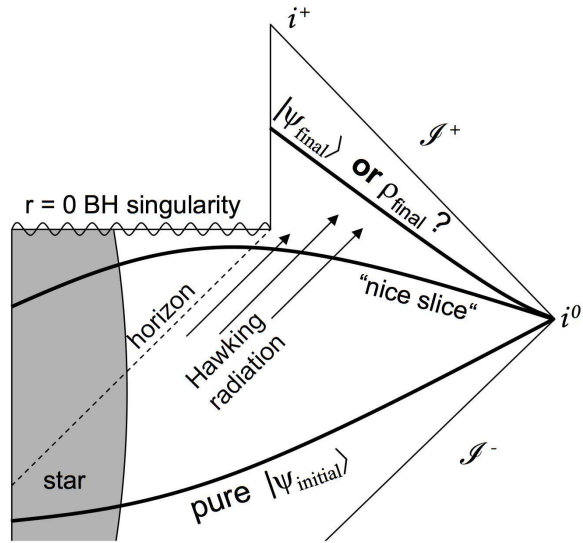}
\end{center}
\caption{\label{bhinfo} Formulation of the black hole information puzzle on a semiclassical spacetime. The no-cloning theorem prevents the same quantum information from existing in two different places on the nice slice. }
\end{figure}

\begin{figure}
\begin{center}
\includegraphics[width=18pc]{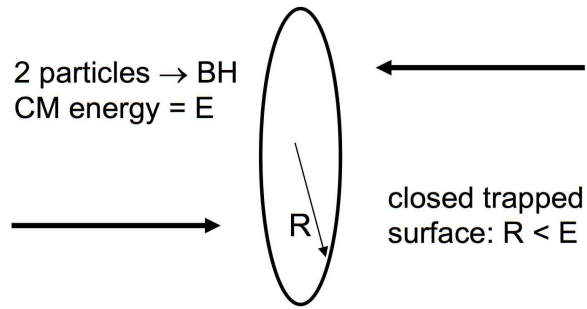}
\end{center}
\caption{\label{bhf} Two particle scattering with sufficiently small impact parameter leads to black hole formation.}
\end{figure}

\section{Decoherence: pure to mixed evolution}

\bigskip
Quantum mechanics as conventionally formulated (the Copenhagen interpretation) allows for two kinds of time evolution: the usual Schr\"odinger evolution, which is unitary, and measurement collapse, which is non-unitary and leads to von Neumann projection onto a particular eigenstate of the operator associated with the measurement. It is appealing to think that wavefunction collapse might only be an apparent phenomenon, which results from unitary evolution. This idea dates to Everett \cite{E}, but has been developed substantially in recent decades as the theory of decoherence \cite{deco}.

\begin{figure}
\begin{center}
\includegraphics[width=18pc]{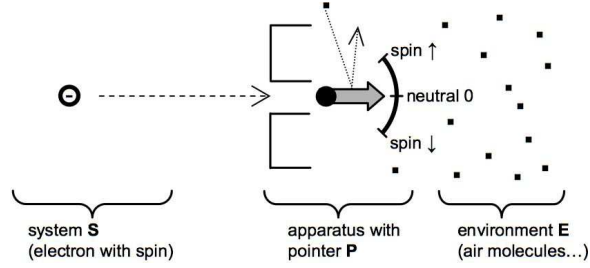}
\end{center}
\caption{\label{meas} Measurement of a single spin leads to entangled state in Eqn.~(\ref{entangle}).}
\end{figure}

Consider a system prepared in a superposition state (see Figure \ref{meas})
\begin{equation}
\vert \psi \rangle = c_1 \vert 1 \rangle + c_2 \vert 2 \rangle~~.
\end{equation}
Suppose that, due to interactions between the system and its environment (or, equivalently, a measuring apparatus), the two evolve into an entangled state
\begin{equation}
\label{entangle}
\vert \psi \rangle \otimes \vert E \rangle ~~\rightarrow~~ \vert \Psi \rangle = c_1 \vert 1 \rangle \otimes \vert E_1 \rangle + c_2 \vert 2 \rangle \otimes \vert E_2 \rangle~~.
\end{equation}
The states $E_{1,2}$ are referred to as pointer states of the environment or measuring device. These pointer states are determined by the dynamics -- that is, the $1,2$ bases along which the device makes measurements is determined by its specific properties. A Stern-Gerlach machine measures spin along a particular axis given by its magnetic field; it will not decohere the system into spin states along any other directions. After the measurement interaction, the state of the system is encoded redundantly in the environment states $E_{1,2}$ and can be read out by making simple measurements on subsets of the degrees of freedom -- i.e., did the red light flash (look for red photons), or did the green light flash?

The environment is assumed to have a large number of degrees of freedom, so that after a relatively short time (determined by the specific dynamics) the states $E_1$ and $E_2$ are nearly orthogonal. The dimensionality of a Hilbert space describing $N$ degrees of freedom is exponential in $N$: for qubits, $d = 2^N$. Two randomly chosen vectors from this space will have overlap $\langle E_1 \vert E_2 \rangle \sim 1/d$, which is tremendously small for any macroscopic environment or measuring device (e.g., $N \sim$ Avogadro's number).  Consequently, interference phenomena between the two ``branches'' of system plus environment are highly suppressed (see below).

Absent the capability to measure the environmental degrees of freedom, it is appropriate to trace over the degrees of freedom of $E$:
\begin{equation}
\label{rho_hat}
\hat{\rho} ~=~ {\rm Tr}_E \vert \Psi \rangle \langle \Psi \vert ~=~  \vert c_1 \vert^2 \, \vert 1 \rangle \langle 1 \vert + \vert c_2 \vert^2 \, \vert 2 \rangle \langle 2 \vert~~;
\end{equation}
this density matrix, from which the outcome of all subsequent measurements on the system alone can be predicted just as well as from the knowledge of the complete state $\vert \Psi \rangle$, is diagonal if one neglects the exponentially small overlap $\langle E_1 \vert E_2 \rangle$. This process then has the appearance of measurement with fundamental wavefunction collapse and probabilistic outcomes, despite purely unitary Schr\"odinger evolution. For All Practical Purposes -- FAPP, as formulated by Bell \cite{Bell} -- decoherence leads to the usual Copenhagen interpretation. 

But a nagging issue remains -- the presence of the other ``branches'' in the pure state $\vert \Psi \rangle$. Are they real? Can they ever be detected experimentally? The off-diagonal elements of the reduced density matrix are suppressed by the small overlap of typical environmental states $E_1, E_2$, but can their effects be measured?

Omn\`es \cite{Omnes} made detailed estimates of the capabilities required to distinguish between decoherence and fundamental collapse, which we will examine in the following section. He found that for macroscopic objects, e.g., containing Avogadro's number of degrees of freedom, any beyond-FAPP device would have to be larger than the visible universe. Hence, Omn\`es asserted, any distinction between the Copenhagen interpretation and unitary wavefunction evolution (leading to Everett branches) for macroscopic objects is untestable, and beyond the realm of scientific inquiry.

The foregoing discussion, in particular Omn\`es' estimate, assumes big environments (e.g., $N\sim$ Avogadro's number). However, decoherence mechanisms are also at work if the number of degrees of freedom $N$ of the environment is smaller, or if the interaction between system and environment is weaker; in these cases, however, decoherence effects are either not as strong (the off-diagonal elements in $\hat{\rho}$ cannot be completely neglected as in (\ref{rho_hat}), i.e.~FAPP does not hold in this case), or happen only over time scales longer than in the strongly-interacting case, respectively. In recent years, this gradual onset of decoherence in controlled environments (also called the ``Quantum-to-Classical transition'') has been the focus of quite a number of laboratory experiments. For example, in \cite{hornberger} the gradual loss of spatial coherence (interference pattern) of fullerene molecules in a slit experiment was observed with increasing pressure of the gas, i.e., with increasing interaction strength (cross section) between fullerenes and gas molecules (environment). Other experiments, e.g.~\cite{haroche}, have verified the gradual loss of coherence in a system (superposition of two states of a Rydberg atom) with increasing number of degrees of freedom of the interacting environment ($N\sim10$ photons in a cavity). On the other hand, one of the big challenges in achieving useful quantum computing \cite{quantum_computer} is to build and control large and scalable quantum systems ($N\sim100$ or more) in which coherence is maintained (possibly via quantum error correction) over the time of the computation.

In the decoherence approach to measurement an initially pure state is later described by a reduced density matrix which, FAPP, represents a mixed state. The black hole information puzzle is often described in similar terms: infalling matter in a pure state is somehow transformed into a mixed state of Hawking radiation. Or, equivalently, if the quantum information in the black hole precursor is not to be found in the outgoing radiation, the radiation is surmised to be in a mixed state. It has been claimed that pure to mixed evolution implies, necessarily, catastrophic consequences, such as violation of energy conservation \cite{BPS} (see 
\cite{related} for additional arguments, for and against this point of view). Potential resolutions of the puzzle in which the information ends up somewhere outside our universe (e.g., involving baby universes or spacetime topology change \cite{TopologyChange}) must still have an effective description in our universe in terms of pure to mixed state evolution. Decoherence provides an example of {\it effective} pure to mixed evolution without catastrophic consequences. Tracing over the environmental states, one loses track of the total energy of the system, but without any resulting catastrophe.

\bigskip
\section{Black hole information experiments}

\bigskip
Below we describe two categories of experiments which test different aspects of theoretical ideas about black hole information. Despite their differences, both require the ability to detect or manipulate macroscopic superpositions. 

\bigskip
{\bf A. Test purity vs decohered mixed state.} A basic goal would be to differentiate between pure and mixed states of the type produced by decoherence. Without this minimal capability one can hardly investigate whether pure states evolve to mixed states, as proposed originally by Hawking. (Was the initial state pure? Is the final state pure or mixed?)

In the black hole context, one could imagine forming the hole from an initial state with at least some degrees of freedom in a superposition (e.g., two spin states). The remaining degrees of freedom can be considered the environment $E$ from our earlier discussion, assuming the dynamics are such that the environment evolves into two different pointer states corresponding to the superposition. This would be the case if, for example, the two spin states had slightly different energy due to a magnetic field provided by the other degrees of freedom.

If, for each $i=1,2$, the initial state $\vert i\rangle$ of the system leads to the final state $\vert \Psi_i \rangle =  \vert i \rangle \otimes \vert E_i \rangle$ for system plus environment, then, starting from the initial state $c_1\vert1\rangle+c_2\vert2\rangle$, two candidates for the final state would be: on the one hand a pure state superposition
\begin{equation}
\label{proposedpurestate}
\vert \Psi \rangle = 
c_1  \vert \Psi_1 \rangle  + c_2 \vert \Psi_2 \rangle~
\end{equation} 
as predicted by unitary evolution, corresponding to the pure density matrix
\begin{equation}
{\hat{\rho} \,}_P = \vert  \Psi \rangle \, \langle \Psi  \vert~~,
\end{equation}
and, on the other hand, a mixed state density matrix
\begin{equation}
\label{proposedmixedstate}
{\hat{\rho} \,}_M ~=~ \vert c_1 \vert^2 ~ \vert \Psi_1 \rangle ~\langle \Psi_1 \vert  ~+~ 
\vert c_2 \vert^2 ~  \vert \Psi_2 \rangle ~\langle \Psi_2 \vert~~,
\end{equation}
predicted by wavefunction collapse. What is required to differentiate between these alternatives? Consider a measurement operator $M$, which, without loss of generality, we might take to be a projector onto some subspace of the Hilbert space. The pure and mixed states can be distinguished if we can detect the difference
\begin{equation}
\label{diff}
{\rm Tr} [ {\hat{\rho} \,}_P \,  M]  -   {\rm Tr} [ {\hat{\rho} \,}_M \, M  ] \,=\,
c_1^* c_2^{} \langle \Psi_1 \vert M \vert \Psi_2 \rangle + 
c_2^* c_1^{} \langle \Psi_2 \vert M \vert \Psi_1 \rangle~~.
\end{equation}
Let us consider two possibilities. If $M$ is a generic operator -- for example, only couples to a limited subset of the degrees of freedom -- then the matrix elements in (\ref{diff}) will be exponentially small. In particular, 
\begin{equation}
\label{small}
\langle \Psi_1 \vert M \vert \Psi_2 \rangle \sim \frac{1}{\rm{dim}\,{\cal H}_{S \otimes E}} \sim \exp( - N)~~,
\end{equation}
where $N$ is the number of degrees of freedom of system plus environment. For a black hole, $N$ is of order its entropy, or area in Planck units. Omn\`es \cite{Omnes} has argued that when $N$ is of order Avogadro's number (e.g., for one gram of ordinary matter), a measurement of this accuracy is impossible {\it in principle}. Based on this, Omn\`es concludes that questions about decoherent branches other than the one observed as an outcome are not scientific. In effect, his calculations purport to extend Bell's FAPP to a statement of principle. Roughly speaking, he argues that the sensitivity of a classical device with $N'$ degrees of freedom only improves as a power of $N'$ (not exponentially with $N'$). Since the precision needed to detect a decohered macroscopic superposition involving $N$ degrees is $\sim \exp( - N)$, as in (\ref{small}), the required $N'$ grows exponentially with $N$. To detect a macroscopic superposition with $N \sim 10^{24}$, Omn\`es concludes, would require $N'$ larger than the number of particles in the visible universe.

In AdS space a concrete proposal \cite{maldacena} has been made  for how to decide between unitary (pure to pure) and non-unitary (pure to mixed) evaporation of a black hole, by measuring whether a certain correlation function drops, over time, either to zero or to the finite value $\exp(-N)$. Interestingly, the required measurement accuracy is essentially that necessary to detect Everett branches of a system with a similar number $N$ of degrees of freedom, as described in the previous paragraph.

On the other hand, for a carefully engineered operator $M$, the amplitude $\langle \Psi_1 \vert M \vert \Psi_2 \rangle$ in (\ref{diff}) can be of order 1/2, even if $\langle \Psi_1 \vert \Psi_2 \rangle=0$. Indeed, the maximum value is obtained when $M$ is a projector onto a macroscopic superposition of the type $\sim  \vert \Psi_1 \rangle  + \vert \Psi_2 \rangle\,$, like $\vert\Psi\rangle$ itself. It is a formidable challenge to perform a measurement of such an operator $M$, presumably even harder than preparing $\vert\Psi\rangle$ itself, and we give a concrete, albeit idealized, example below in the context of the Coleman-Hepp model of measurement \cite{Hepp} to illustrate the difficulties arising even in simple cases.

Coleman and Hepp proposed an explicit model of a quantum measurement which results from the interaction of spin states. In their model the interaction between the system (itself a spin) and measuring device causes evolution as in (\ref{entangle}), with
\begin{eqnarray}
\label{CH}
\vert \Psi_1 \rangle &=& \vert 1 \rangle \otimes \vert E_1 \rangle = \vert + \rangle \otimes \vert + + + \cdots + \rangle \\
\vert \Psi_2 \rangle &=& \vert 2 \rangle \otimes \vert E_2 \rangle = \vert - \rangle \otimes \vert - - - \cdots - \rangle~~\nonumber,
\end{eqnarray}
where $\pm$ are spin-up and spin-down states along the $z$ axis, and the measuring device has $N$ degrees of freedom (qubits). That is, the interaction between system and measuring device leads to a correlation between the initial spin state and the (macroscopic, when $N$ is large) state of the device. The state of the spin can be read out by measuring some subset of the $N$ degrees of freedom in the device.

In this context it is straightforward to design an operator $M$ for which  $\langle \Psi_1 \vert M \vert \Psi_2 \rangle$ is large: we simply take the tensor product operator
\begin{equation}
M ~=~ \bigotimes_i\,\sigma_x^i~~,
\end{equation}
where $\sigma_x^i \vert \pm \rangle = \vert \mp \rangle$ measures the spin of environmental qubit $i$ in the $x$-direction. An experimental realization of $M$ would be able to distinguish the two considered post-measurement states of the system plus device: pure ${\hat{\rho} \,}_P$ vs.~mixed ${\hat{\rho} \,}_M$. Note, though, that the simple $M$ considered in this toy example is not a projection (in particular, not the projector into the one-dimensional subspace $\vert\Psi\rangle$) and so cannot, e.g., distinguish the considered macroscopic superposition pure state (\ref{proposedpurestate}) from the (seemingly simpler) pure state $\vert +_x \rangle \otimes \vert +_x +_x\cdots +_x\rangle$ in which all spins are aligned in the $+x$-direction.

One might object that in this simple example the pointer states $E_{1,2}$ of the device in (\ref{CH}) are mutually orthogonal, but not thermalized. In this sense the model does not represent a realistic measurement (the ``environment'' is highly constrained). This could easily be remedied by allowing some interactions between the spins in the device, which leads to some (presumably) ergodic but unitary evolution. If one keeps the spins isolated from the rest of the universe, these interactions evolve $\Psi_{1,2}$ into something more random, at least in appearance: 
\begin{equation}
\vert\Psi\rangle~\to~\vert \Psi' \rangle ~=~ U \, ( c_1 \vert \Psi_1 \rangle + c_2 \vert \Psi_2 \rangle )
\end{equation}
or, for the proposed mixed state (\ref{proposedmixedstate}) after the measurement,
\begin{equation}
{\hat{\rho} \,}_M~\to~{\hat{\rho}'}_M~=~U{\hat{\rho} \,}_MU^\dagger~~.
\end{equation}
It would still be the case that the operator $M' = U M U^\dagger$ can distinguish between pure and mixed states of the post-measurement device. However, if the $N$ spins are separated in space (e.g., correspond to isolated qubits in a quantum device), then the operator $M'$ would itself have to be realized out of macroscopic superpositions of spatially separated objects, unlike the original $M$ which acted independently on each of the spins. 

To summarize, a measurement which can differentiate between pure and mixed states either has to rely on extreme precision to detect very small matrix elements as in (\ref{small}), or on the ability to prepare a very special, typically non-local, operator like $M'$.

\bigskip
{\bf B. Test Hawking mixed state vs typical pure state.} In this scenario we compare Hawking radiation in a mixed state $\rho_*$ to radiation in a pure state $\psi$. 

It is widely believed that large scale violation of locality at the black hole horizon is necessary for unitary evolution. This might lead to significant deviations from Hawking's results describing what is emitted from the hole. If such deviations were observed, they would undermine the usual semiclassical reasoning which leads to the information puzzle, although deviations of the radiation from Hawking's mixed state description do not by themselves imply that evolution is unitary or purity-conserving.

Perhaps a more plausible scenario, assumed in what follows, is that the radiation, although described by a pure state $\psi$, only deviates in subtle ways from the Hawking mixed state. That is, information is encoded in correlations (phases or superpositions) between particle states in the radiation, but the overall distribution appears to be thermal and the temperature evolution is as predicted by Hawking. 

It is extremely difficult to differentiate between a pure state $\psi$ of this type and the Hawking mixed state $\rho_*$. Local measurements on the radiation (i.e., over length scales much smaller than its full spatial extent) can only exclude tiny subsets of the $\psi$ Hilbert space. Moreover, it can be shown that for almost all states $\psi$ these local measurements are governed by the same (thermal) probability distribution as the one obtained from $\rho_*$.

\begin{figure}[h]
\begin{center}
\includegraphics[width=18pc]{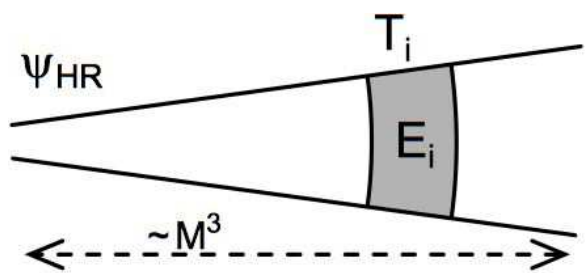}
\end{center}
\caption{\label{hr} Hawking radiation emitted in a pure state $\psi_{HR}$ which conforms to the semiclassical prediction for temperature and energy radiated in each time interval $i$.}
\end{figure}

A simple way to understand this is to recall that maximizing the entropy of a system subject to an energy constraint leads to a thermal distribution. Pure states which conform, at least macroscopically, to the Hawking predictions are constrained to describe the same total energy emission over any particular interval of time. To be specific, consider a time interval $\Delta t_i$ over which the Hawking temperature is close to constant, $T = T_i$, but during which many particles are emitted. Let the Hawking calculation predict that a total amount of energy ${\cal E}_i$ be emitted during the interval (see Figure \ref{hr}). Ordinary statistical mechanics tells us that the overwhelming majority of states (of the system in the volume corresponding to $\Delta t_i$) with energies close to ${\cal E}_i$ will be approximately thermal  -- i.e., maximizing the entropy leads to a Boltzmann distribution for energy occupation numbers, with temperature equal to the Hawking temperature $T_i$. The probability distribution governing measurements of the energy distribution of individual emitted quanta over the interval $\Delta t_i$ will then coincide with that given by $\rho_*$, except for an exponentially rare subset of states satisfying the constraint (i.e., configurations with much lower entropies than the Boltzmannian, or thermal, ones).

A more explicit computation follows. Consider the subset of pure states $\psi$ which conform to the Hawking predictions governing the amount of energy radiated in each time interval $\Delta t_i$. Specifically, require that, for every $i$, the reduction (by taking the partial trace) of $\psi$ to the degrees of freedom emitted in the interval $\Delta t_i$ be a mixture of superpositions of the energy eigenstates (of the theory in that volume) with eigenvalues close to ${\cal E}_i$. That is, consider region $i$ and only the degrees of freedom within it (neglecting boundary effects, which are negligible for large regions). The reduction of $\psi$ to these degrees of freedom, when expanded in an energy eigenstate basis for the region $\Delta t_i$, must have support only on states with energies close to ${\cal E}_i$. The superposition of two pure states $\psi_1$ and $\psi_2$ satisfying this condition will also satisfy the condition as superposition does not extend the region of support; therefore the condition defines a subspace of the larger Hilbert space. Denote by ${\cal H}_R$ this restricted (``energetically conforming'') subspace of the overall radiation Hilbert space $\cal H$.

Further, divide the radiation into a subsystem $S$, to be measured, and the remaining degrees of freedom which constitute an environment $E$, so ${\cal H} = {\cal H}_S \otimes {\cal H}_E$ and 
 \begin{equation}
 \rho_S \equiv\rho_S(\psi)= {\rm Tr}_E \vert \psi \rangle \langle \psi \vert
\end{equation}
is the density matrix which governs measurements on $S$ for a given pure state $\psi$. Note the assumption that these measurements are local to $S$, hence the trace over $E$.

Then, a recent theorem \cite{Winters} on entangled states, which exploits properties of Hilbert spaces of very high dimension, shows that for almost all $\psi \in {\cal H}_R$, $\rho_S(\psi) \approx {\rm Tr}_E\left(\rho_*\right)$ as long as $d_E \gg d_S$, where $d_{E,S}$ are the dimensionalities of  the ${\cal H}_E$ and ${\cal H}_S$ Hilbert spaces. In the theorem, $\rho_* = {1}_R / d_R$ is the equiprobable mixed state on the restricted Hilbert space ${\cal H}_R$ (${1}_R$ is the identity projection on ${\cal H}_R$ and $d_R$ the dimensionality of ${\cal H}_R$), so ${\rm Tr}_E\left(\rho_*\right)$ is the corresponding canonical state of the subsystem $S$. In other words, $\rho_*$ describes a perfectly thermalized radiation system with temperature profile equal to that of  Hawking radiation from an evaporating black hole.

To state the theorem in \cite{Winters} more precisely, the (measurement-theoretic) notion of the \emph{trace-norm} is required, which can be used to characterize the distance between two mixed states $\rho_S$ and $\Omega_S$:
\begin{equation}
\Vert\rho_S-\Omega_S\Vert_1\equiv{\rm Tr}\sqrt{\left(\rho_S-\Omega_S\right)^2}~~.
\end{equation}
This sensibly quantifies how easily the two states can be distinguished by measurements, according to the identity
\begin{equation}
\label{tracesup}
\Vert\rho_S-\Omega_S\Vert_1 = {\rm sup}_{\Vert O\Vert\leq1}\,{\rm Tr}\left(\rho_SO-\Omega_SO\right)~~,
\end{equation}
where the supremum runs over all observables $O$ with operator norm $\Vert O\Vert$ smaller than 1 (projectors $P=O$ are in some sense the best observables, all other observables can be composed out of them, and they have $\Vert P\Vert=1$). Note that the trace on the right-hand side of (\ref{tracesup}) is the difference of the observable averages $\langle O\rangle$ evaluated on the two states $\rho_S$ and $\Omega_S$, and therefore specifies the experimental accuracy necessary to distinguish these states in measurements of $O$. A special form of the theorem then states that the probability that
\begin{equation}
\Vert\rho_S\left(\psi\right)-{\rm Tr}_E\left(\rho_*\right)\Vert_1 ~\geq~ d_R^{-1/3}+\sqrt{\frac{d_S^{\,2}}{d_R}}
\end{equation}
is less than $2 \exp ( -d_R^{1/3}/18\pi^3 )$. In words: let $\psi$ be chosen randomly (according to the natural Hilbert space measure) out of the space of allowed states ${\cal H}_R$; 
the probability that a measurement on the subsystem $S$ \emph{only}, with measurement accuracy of $d_R^{-1/3}+\sqrt{d_S^{\,2}/d_R}$, 
will be able to tell the pure state $\psi$ (of the entire system) apart from the mixed state $\rho_*$ is exponentially small ($\sim\exp -d_R^{1/3}$) in the dimension of the space ${\cal H}_R$ of allowed states. 
That is, the overwhelming majority of energetically conforming pure Hawking evaporation states $\psi \in{\cal H}_R$ cannot be distinguished from Hawking's predicted density matrix $\rho_*$ by measurements on a small subsystem $S$ of the radiation, even if the experimental error in measurements of projectors is only $d_R^{-1/3}+\sqrt{d_S^{\,2}/d_R}$. This is an incredible precision, considering the estimates $d_R\sim\exp S_{BH}\sim\exp M^2$ and $d_S\sim\exp S_S$, where the entropy $S_S\sim VT^3\ll S_{BH}$ of the (energetically conforming) subsystem $S$ can be computed from its volume $V$ and the Hawking temperature $T$ of the particle excitation distribution inside. 

Thus, as long as individual measurements are localized in spacetime, so that $S$ is small relative to $E$, one cannot distinguish a typical state $\psi \in {\cal H}_R$ from $\rho_*$ without exponential sensitivity in the measurement on $S$. This is true even if one performs many measurements on distinct subsystems $S_i$, in particular even if the union of $S_i$ covers all of the radiation. This is because, even if each subsystem were in a pure state that could in principle be measured exactly (as opposed to merely measuring all the single-particle excitations in it separately), at the very least the phase relations between the states of the different subsystems $S_i$ are lost. Only complete measurements (including phase relations) on very big subsystems $S$ ($\sim$ half of the degrees of freedom of $E$) have a non-infinitesimal chance of distinguishing $\psi$ from $\rho_*$.

In analogy to what we learned in the previous cases, measurements that can distinguish generic $\psi \in {\cal H}_R$ from $\rho_*$ with reasonable probability, or without exquisite sensitivity, must be highly non-local, covering a spacetime region which includes most of the radiation emitted by the black hole. They must, in a sense, measure it all ``at once''. But measurements of this sophistication could, again, also differentiate between macroscopic superpositions and mixed states.

\bigskip
\section{Conclusions}

\bigskip
Black hole information experiments are at least as hard as experiments which test decoherence vs fundamental collapse. One has to create a semiclassical black hole (in fact, many of them in \emph{identical} states) and then make very challenging measurements on the relativistic decay products, which include gravitons. (Note, it appears difficult to determine the quantum state of gravitons with physically realizable detectors that obey the positive energy condition \cite{gravitons}.) In particular, the experiment must be sensitive to the phase relations in coherent superpositions of Fock states rather than simply counting occupation numbers as ordinary particle detectors do. By comparison, the most accessible tests of decoherence would be in the context of an artificial toy system like the Coleman-Hepp model, or other controlled quantum computing environment.

Our results can be summarized rather simply. Hawking suggested black holes cause pure states to evolve to mixed states. But, for all practical purposes (FAPP), decoherence does the same thing -- {\it or at least appears to}. In order to test Hawking's proposal one therefore has to go beyond FAPP and beyond decoherence. Such capability allows fundamental tests of quantum measurement.

If fundamental questions about measurement, decoherence and wavefunction collapse are philosophy rather than science -- i.e., cannot be tested by experiments -- then so is the black hole information puzzle.

\bigskip
\emph{Acknowledgments ---} The author thanks the organizers of the MCCQG for a stimulating and pleasant meeting. This work was supported by the Department of Energy under DE-FG02-96ER40969.



\begin{thebibliography}{99}
\bibitem{scatter} In analogy with ordinary scattering theory we could write the quantum state at future infinity as $\psi' = \psi_0 + \psi_1$ where $\psi_0$ is the future-evolved unscattered state (i.e., no black hole formed; initial state particles can propagate to infinity) and $\psi_1$ the Hawking radiation state (black hole was formed and evaporated). For suitable initial data (e.g., a large ball of dust) one can ensure that the magnitude of  $\psi_0$ is arbitrarily small (a black hole is almost certainly formed) and that $\psi_1$ is largely independent of the internal state of the hole (the no-cloning argument applies to the semiclassical black hole spacetime). This suggests a unitarity problem: if varying the internal state does not change $\psi_1$, then many distinct states are mapped to almost the same final state $\psi' = \psi_0 + \psi_1 \approx \psi_1$. This argument does not, however, require the final state to be a mixture.
\end{thebibliography}
\end{document}